\documentclass[12pt]{article} 
\usepackage{amsmath,amsthm,amscd,amsfonts,amssymb}
\usepackage{latexsym}
\newcommand{\CC}{\mathbb C}

\newcommand{\RR}{\mathbb R}
\newcommand{\TT}{\mathbb T}
\newcommand{\ZZ}{\mathbb Z}
\newcommand{\EE}{\mathbb E}
\newcommand{\Znu}{\ZZ^\nu}
\newcommand{\Znuplus}{\ZZ^{\nu +1}_+}
\newcommand{\cc}{\mathcal C}
\newcommand{\dd}{\mathcal D}
\newcommand{\hh}{\mathcal H}
\newcommand{\ww}{\mathcal W}
\newcommand{\WW}{\mathcal S}
\newtheorem{thm}{Theorem}[section]
\newtheorem{lemma}[thm]{Lemma}
\newtheorem{cor}[thm]{Corollary}

\newtheorem{defin}[thm]{Definition}

\newtheorem{hyp}[thm]{Hypothesis}
\def\thet{{\bf \vartheta}}
\begin{document}
\title{Spectra of Anderson type models with decaying randomness}
\author{M Krishna \\ 
Institute of Mathematical Sciences\\ Taramani,
Chennai 600 113, India \\
K B Sinha\\
Indian Statistical Institute\\
7 S J S Sansanwal Marg, New Delhi 110016}\footnotetext{This work is 
supported by the grant DST/INT/US(NSF-RP014)/98 of the Department of Science and
Technology }
\date{}
\maketitle
\begin{abstract}
In this paper we consider some Anderson type models, with 
free parts having long range tails and with the
random perturbations decaying at different rates in different
directions and prove that there is a.c.
spectrum in the model which is pure.  In addition, we show that there
is pure point spectrum outside some interval.  Our models include
potentials decaying in all directions in which case absence of singular 
continuous spectrum is also shown. 
\end{abstract}

\section{Introduction}

There have been but few models in higher dimensional random operators
of the Anderson model type in which presence of absolutely continuous
spectrum is exhibited.  We present here one family of models with
such behaviour.

The results here extend those of Krishna \cite{mk1} and
part of those in Kirsch-Krishna-Obermeit \cite{kko}, Krishna-Obermeit
\cite{ko} while making use of wave operators to show the existence of
absolutely continuous spectrum, 
the results of Jaksic-Last \cite{jl} to show its purity 
and those of Aizenman \cite{ma} for exhibiting pure point spectrum.

Our models include the independent randomness on a surface  
considered by Jaksic-Molchanov \cite{jm, jm2} and Jaksic-Last
\cite{jl}, \cite{jl2}, while allowing for the randomness to extend
into the bulk of the material. 

The literature on the scattering theoretic and commutator methods for discrete
Laplacian includes those of Boutet de Monvel-Sahbani \cite{ms1},
\cite{ms2} who study deterministic operators on the lattice.

The scattering theoretic method that we use is  applicable even
when the free operator is not the discrete Laplacian but has 
long range off diagonal parts.  We impose conditions on the free part
in terms of the structure it has in its spectral representation.

\section{Main results }

In this section we present the model we consider in this paper.
We consider the discrete Laplacian on $\ell^2(\ZZ^\nu)$
and the discrete Laplacian $(\Delta u)(n) = \sum_{|i| = 1} u(n+i)$,
which is unitarily equivalent to the operator of multiplication by the function 
$h(\thet) = 2 \sum_{j=1}^\nu \cos(\theta_i)$ acting on $L^2(\TT^\nu,
d\sigma(\thet))$, where $\TT = [0, 2\pi]$ and $d\sigma(\thet) =
\prod_{i=1}^\nu \frac{d\theta_i}{2\pi}$, where we use the
notation $\thet = (\theta_1, \cdots, \theta_\nu)$.  We consider a
bounded self adjoint operator $H_0$ which commutes with $\Delta$ and
which is given by, in $L^2(\TT^\nu, d\sigma)$, an operator of
multiplication by a function $h(\thet)$ there with h satisfying the
assumptions below.

\begin{hyp}  
\label{hyp1.1}
Let h be a real valued $C^{3\nu+3}(\TT^\nu)$ function satisfying 
\begin{enumerate}
\item h is separable, i.e. $h(\thet) = \sum_{j=1}^\nu h_j(\theta_j)$.
\item The sets 
$$
\cc(h_j) = \{ x : \frac{d h_j}{d\theta}(x) =0 \}
$$
are finite for each $j = 1, \cdots, \nu$.  We denote by
$$
\cc = \cup_{j=1}^\nu \cc(h_j)
$$
and note that this is a set of measure zero in $\TT^\nu$.
\item Each of the $h_j$ and its derivatives, up to the highest order, 
satisfy $h_j^{(k)}(0) = h_j^{(k)}(2\pi), j = 1, \cdots, \nu, ~~ k = 0,
\cdots, 3\nu + 3$.

\end{enumerate}
\end{hyp}
{\noindent \bf Remark:} The values of the functions $h_i$ and its 
derivatives at the boundary points is
interpreted as $h^{(j)}(0) = \lim_{\epsilon \downarrow 0}
h^{(j)}(\epsilon)$ and similarly $h^{(j)}(2\pi) = \lim_{\epsilon
\downarrow 0} h^{(j)}(2\pi -\epsilon)$.  

We consider random perturbations of bounded self adjoint operators
coming from functions as in the above hypothesis.  We assume the
following on the distribution of the randomness.

\begin{hyp}
\label{hyp1.2}
Let $\mu$ be a positive probability measure on $\RR$ satisfying:
\begin{enumerate}
\item $\mu$ has finite variance $\sigma^2 = \int x^2 d\mu(x) $.
\item $\mu$ is absolutely continuous.
\end{enumerate}
\end{hyp}

Finally we consider some sequences of numbers $a_n$ indexed by the
lattice $\Znu$ or $\ZZ^{\nu +1}_+ = \ZZ^+ \times \Znu$ and assume the 
following on them.

\begin{hyp}
\label{hyp1.3}
\begin{enumerate}
\item $a_n$ is a bounded sequence of non-negative numbers  
 indexed by $\Znu$ which is non-zero on an infinite subset of $\Znu$. 
\item  Let $g(R) = a_n \chi_{\{n \in \ZZ^\nu : |n_i| > R, ~~ \forall 
1 \leq i \leq \nu\}}$.  Then $g \in L^1((1, \infty))$ 
\item $a_n$ is a bounded sequence of non-negative numbers which are
non-zero on an infinite subset of $\ZZ^{\nu + 1}_+$.
\item  Let $g(R) = a_n \chi_{\{n \in \ZZ^\nu : |n_i| > R, ~~ \forall   
1\leq i\leq \nu +1\}}$.  Then $g \in L^1((1, \infty))$ 
\end{enumerate}
\end{hyp}

{\noindent \bf Remark:} 1. In the case of $\ZZ^\nu$ our hypothesis on
the sequence $a_n$ allows for the following type of sequences
\begin{itemize}
\item $a_n = (1+|n|)^\alpha, ~~ \alpha < -1$.
\item $a_n = (1+ |n_i|)^\alpha, \text{for some } ~ i, ~~ \alpha < -1$.
\item $a_n = \prod_{i=1}^\nu (1+ |n_i|)^{\alpha_i}, ~~ \alpha_i \leq 0
~~ \text{with}~~ \sum_{i=1}^\nu \alpha_i < -1$.
\end{itemize}
Therefore in the theorems, on the existence of absolutely continuous
spectrum, we can allow the potentials to be stationary
along all but one direction in dimensions $\nu \geq 2$.

2. In the case of $\ZZ^{\nu+1}_+$, we can allow the sequence to be of
the type
\begin{itemize}
\item $a_n = 0, ~~ n_1 > 0 ~~ \text{and} ~~ a_n = 1, ~~\text{for} ~~ n_1 = 0$.
\item $a_n = (1+ |n_1|)^\alpha, ~~ \alpha < -1$.
\item $a_n = \prod_{i=1}^\nu (1+ |n_i|)^{\alpha_i}, ~~ \alpha_i \leq 0
~~\text{with} ~~  \sum_{i=1}^\nu \alpha_i < -1$.
\end{itemize}
Thus allowing for models with randomness on a the boundary of a half space.

For the purposes of determining the spectra of the models we are going
to consider here in this paper we recall a definition given in 
Kirsch-Krishna-Obermeit \cite{kko}, namely,

\begin{defin}
\label{def1.1}
Let $a_n$ be a non-negative sequence, indexed by $\Znu$ or $\Znuplus$.  
Let
$\mu$ be a positive probability measure on $\RR$.  Then the {\bf
a-supp}($\mu$) is defined as 
\begin{enumerate}
\item In the case of $\Znu$,
$$
\text{a-supp}(\mu) = \cap_{k \in \ZZ^+, k\neq 0}
\{x : \sum_{n\in k\Znu} \mu(a_n^{-1}(x - \epsilon,
x+\epsilon)) = \infty, ~~ \forall ~~ \epsilon > 0\}.
$$
\item In the case of $\Znuplus$,
$$
\text{a-supp}(\mu) = \cap_{k \in \ZZ^+, k\neq 0}
\{x : \sum_{n\in k\Znuplus} \mu(a_n^{-1}(x - \epsilon,
x+\epsilon)) = \infty, ~~ \forall ~~ \epsilon > 0\}.
$$
\end{enumerate}
\end{defin}

{\noindent \bf Remark:} 1. In the sums occurring in the above
definition we set $\mu(a_n^{-1}(x-\epsilon, x+\epsilon)) \equiv 0$,
notationally, for those n for which $a_n =0$.  This notation is to
allow for sequences $a_n$ that are everywhere zero except on an axis
for example. 

2. We note that when $a_n$ is a constant sequence
$a_n = \lambda \neq 0$, 
$$
a-supp(\mu) = \lambda \cdot \text{supp}(\mu).  
$$

3. When $a_n$ converge to zero as $|n|$ goes to $\infty$ then the
$a-supp(\mu)$ is trivial $\mu$ has compact support.  It could be
trivial even for some class of $\mu$ of infinite support depending
upon the sequence $a_n$.

4. If $a_n$ is bounded below by a positive number on an infinite
subset along the directions of the axes in $\Znu$ (respectively
$\Znuplus$), then the $a-supp(\mu)$ could be non-trivial even for
compactly supported $\mu$.

We consider either $\Znu$ or $\ZZ^{\nu+1}_+$
consider the operator, defined by (for $u \in \ell^2(\ZZ^+)$), 
\begin{equation*}
(\Delta_+ u)(n) = 
\begin{cases}
& u(n+1) + u(n-1), ~~ n > 0,\\
& u(1), ~~ n = 0. ~~ 
\end{cases}
\end{equation*}
Below we use either $\Delta_+$ or its extension to $\ell^2(\Znuplus)$
by $\Delta_+\otimes I$, but we abuse notation and continue to denote
this and its extension by $\Delta_+$ 
, the correct operator is understood from the context.
We consider the, bounded self adjoint operators on $\ell^2(\Znu)$
coming from functions h as in hypothesis (\ref{hyp1.1}) given by
$$
(H_0 u)(n) = \int_{\TT^\nu} d\sigma(\vartheta)~~ h(\vartheta)
e^{-i(n\cdot\vartheta)}\widehat{u}(\vartheta), 
$$
where $\widehat{u}$ is the function in $L^2(\TT^\nu, \sigma )$ with $u(n)$ as its
Fourier coefficients, 
$\sigma$ is the normalized invariant measure on $\TT^\nu$, with
$\TT^\nu = [0, 2\pi]^{\nu}$, where $[0, 2\pi]$ (is also identified with
the unit circle as and when necessary and is understood from the context)
We also denote the extension of $I\otimes H_0$ to $\ell^2(\Znuplus)$
by the symbol $H_0$ and $L^2(\TT^\nu, \sigma)$ as simply $L^2(\TT^\nu)$
in the sequel.

We then consider the random operators 

\begin{equation}
\label{themodel}
\begin{split}
H^\omega &= H_0 + V^\omega, ~~ V^\omega = \sum_{n\in I} a_n
q^\omega(n) P_n, ~~\text{on}~~ \ell^2(\Znu),\\  
H^\omega_+ &=  H_{0+} + V^\omega, ~~ V^\omega = \sum_{n\in I} a_n
q^\omega(n) P_n, H_{0+} = \Delta_+ + H_0, ~~\text{on}~~
\ell^2(\Znuplus).\\  
\end{split}
\end{equation}
where $P_n$ is the orthogonal projection onto the one dimensional
subspace generated by $\delta_n$ when $\{\delta_n\}$ is the standard basis 
for $\ell^2(I)$ (I = $\Znu$ or $\Znuplus$).  $q^\omega(n)$ are independent 
and identically distributed real valued random variables with distribution $\mu$.
The operator $H_0$ is some bounded self adjoint operator to be
specified in the theorems later.

Then our main theorems are the following.  First we state a general
theorem on the spectrum of $H_0$ in such models.
For this we consider the operator $H_0$
to denote a bounded self adjoint operator on $\ell^2(\Znu)$ coming from
a function h satisfying the hypothesis \ref{hyp1.1} and $\Delta_+$
defined as before.

\begin{thm}
\label{thm1.0}
Let $H_0$ and $H_{0+}$ be the operators defined as in equation
(\ref{themodel}), coming from functions h satisfying the Hypothesis
(\ref{hyp1.1})(1)(2).  Let 
$$
E_+ = \sum_{j=1}^\nu \sup_{\theta \in [0, 2\pi]} h_i(\theta), ~~
E_- = \sum_{j=1}^\nu \inf_{\theta \in [0, 2\pi]} h_i(\theta).
$$ 
Then, the spectra of both $H_0$ and $H_{0+}$ are purely absolutely
continuous and
$$
\sigma(H_0) = [E_-, E_+], ~~\text{and} ~~ \sigma(H_{0+}) \supset [-2 +
E_-, 2 + E_+].
$$
\end{thm}

Part of the essential spectra of the operators $h^\omega$ and
$H^\omega_+$ are determined
via Weyl sequences constructed from rank one perturbations of the free
operators $H_0$ and $H_{0+}$ respectively.  The proof of this theorem
is done essentially on the line of the proof of theorem 2.4 in
\cite{kko}.  

\begin{thm}
\label{thm1.1}
Let the indexing set I be $\Znu$ or $\ZZ^{\nu+1}_+$ and consider the
operator $H_0$ coming from a function h satisfying the conditions of
Hypothesis \ref{hyp1.1}(1) in the case of $I = \Znu$ and consider the
associated $H_{0+}$ in the case of $\Znuplus$ the case of  $I = \Znuplus$. Suppose
$q^\omega(n), ~~ n\in I$ are i.i.d random variables with the distribution $\mu$
satisfying the hypothesis \ref{hyp1.2}(1). Let $a_n$ be a sequence
indexed by I satisfying the Hypothesis \ref{hyp1.3}(1) (or (3) as the
case may be).  Assume also that $0 \in a-supp(\mu)$, then 
$$
\bigcup_{\lambda \in ~~ a-supp(\mu)} \sigma(H_0 +\lambda P_0) 
\subset \sigma(H^\omega)
 ~~ \text{almost every} ~~ \omega.
$$
and
$$
\bigcup_{\lambda \in ~~ a-supp(\mu)} \sigma(H_{0+} +\lambda P_0) 
\subset \sigma(H^\omega_+)
 ~~ \text{almost every} ~~ \omega.
$$

\end{thm}

{\noindent \bf Remark:} 1. When $\mu$ has compact support and $a_n$
goes to zero at infinity, or when $\mu$ has infinite support but $a_n$
has has appropriate decay at infinity, there is no essential spectrum outside that
of $H_0$ for $H^\omega$ almost every $\omega$. So the point of this
theorem is to show that there is essential spectrum outside that of 
$H_0$ based on the properties of the pairs ($a_n$, $\mu$).
  
2. In Kirsch-Krishna-Obermeit \cite{kko} some  
examples of random potentials which have essential spectrum
outside $\sigma(H_0)$ even when $a_n$ goes to zero at $\infty$ were
given.  The examples presented there had $a-supp(\mu)$ as a half axis or the
whole axis, this is because of the decay of the sequences $a_n$.  Here
however, since we allow for $a_n$ to be constant along some
directions, our examples include cases where the spectra of $H^\omega$
are compact with some essential spectrum outside $\sigma(H_0)$.  

We let $E_{\pm}$ be as in theorem (\ref{thm1.0}).  We also set
$\hh_{\omega, n}$ to be the cyclic subspace generated by $\delta_n$
and $H^\omega$.

\begin{thm}
\label{thm1.2}
Consider a bounded self adjoint operator $H_0$ coming from a function
h satisfying the conditions of Hypothesis \ref{hyp1.1}(1)-(3).  Suppose
$q^\omega$ are i.i.d random variables with the distribution $\mu$
satisfying the hypothesis \ref{hyp1.2}(1). 
\begin{enumerate}
\item Let $I = \Znu$ and $a_n$ be a sequence
satisfying the Hypothesis \ref{hyp1.3}(1)-(2).
Then,
$$
\sigma_{ac}(H^\omega) \supset [E_-, ~ E_+] ~~ \text{almost every} ~~
\omega.
$$
Further when $\mu$ satisfies the Hypothesis \ref{hyp1.3}(2), $a_n \neq
0$ on $\Znu$, $\hh_{\omega, n}$, $\hh_{\omega, m}$ not mutually
orthogonal for any n, m in $\Znu$ for almost all $\omega$ and
$E_\pm$ as in theorem (\ref{thm1.0}), we also have
$$ 
\sigma_s(H^\omega) \subset \RR \setminus (E_-,~ E_+) ~~ \text{almost
every} ~~ \omega. 
$$
\item Let $I = \Znuplus$ and $a_n$ be a sequence
satisfying the Hypothesis \ref{hyp1.3}(3)-(4).
Then,
$$
\sigma_{ac}(H^\omega_+) \supset [-2+E_-,~ 2+E_+] 
~~ \text{almost every} ~~ \omega.
$$
Further when $\mu$ satisfies the Hypothesis \ref{hyp1.3}(2), 
$a_n \neq 0$ on a subset of $\Znuplus$ that contains the surface
$\{(0,n): n \in \Znu\}$, the subspaces $\hh_{\omega, n}$,
$\hh_{\omega, m}$ are not mutually orthogonal almost every $\omega$
for m, n in $\{(0, k): k \in \Znu\}$,   
we also have
$$ 
\sigma_s(H^\omega) \subset \RR \setminus (-2, +E_-,~ 2+E_+) ~~ \text{almost
every} ~~ \omega. 
$$
\end{enumerate}
\end{thm}

{\noindent \bf Remark:} 1. When $\mu$ is absolutely
continuous the theorem says that the spectrum of $H^\omega$ in $(E_-,
E_+)$ (respectively in $(-2+E_-, 2+E_+)$ for the $\Znuplus$ case)
is purely absolutely continuous, this is a consequence of a
remarkable theorem of Jaksic-Last \cite{jl} who showed that in such
models with independent randomness, with the randomness non-zero
a.e. on a sufficiently big set ($H_0$ can be any bounded self 
adjoint operator in their theorem, provided the set of points where
the randomness lives gives a cyclic family for the operators
$H^\omega$), whenever there is an interval of a.c. spectrum it is pure 
almost every $\omega$.  Their proof is based on considering spectral measures
associated with rank one perturbations and comparing the spectral
measures of different vectors (which give rise to the rank one
perturbations).

2. Our theorem extends the models of
surface randomness considered by Jaksic - Last \cite{jl2}, to allow
for thick surfaces where the randomness is located in a strip beyond
the surface into the bulk of the material.  Such models are obtained
by taking $a_n = 0, ~ n_1 > N$ for some finite N.  Then such models 
also have their  
spectrum in $(-2\nu-2, 2\nu+2)$ purely absolutely continuous.  The
purity is again a  consequence of a theorem of Jaksic-Last \cite{jl}.

Finally we have the following theorem on the purity of a part of the pure point
spectrum.  We denote 
\begin{equation}
\label{eqn1.0}
\begin{split}
e_+ &= \sup \sigma(H_{0+}), ~~ e_- = \inf \sigma(H_{0+}) ~~ and ~~ e_0 = max (|e_-|,
|e_+|) \\
\end{split}
\end{equation}

\begin{thm}
\label{thm1.3}
Consider a bounded self adjoint operator $H_0$ coming from a function
h satisfying the conditions of Hypothesis \ref{hyp1.1}(1)-(3).  Let I
be the indexing set and suppose 
$q^\omega(n), ~~ n \in I$ are i.i.d random variables with the distribution $\mu$
satisfying the hypothesis \ref{hyp1.2}(1)-(2).  Assume further that
the density $f(x) = d\mu(x)/dx$ is bounded.  Set $\sigma_1 = \int
d\mu(x) |x|$.  Then, 
\begin{enumerate}
\item Let $I = \Znu$ and let $a_n$ be a sequence
satisfying the Hypothesis \ref{hyp1.3}(1)-(2). 
Then there is a critical energy $E(\mu) > E_0$ depending upon the
measure $\mu$ such that
$$
\sigma_{c}(H^\omega) \subset (-E(\mu), E(\mu) ) ~~ 
\text{almost every} ~~
\omega.
$$
\item Let $I = \Znuplus$ and let $a_n$ be a sequence
satisfying the Hypothesis \ref{hyp1.3}(3)-(4). 
Then there is a critical energy $e(\mu) > e_0$ such that
$$
\sigma_{c}(H^\omega_+) \subset (-e(\mu), e(\mu) ) ~~ 
\text{almost every} ~~ \omega.
$$
\end{enumerate}
\end{thm}

{\noindent \bf Remark:} 1. The $E(\mu)$ and $e(\mu)$, while finite
may fall outside the spectra of the operators $H^\omega$ and
$H^\omega_{+}$, for some pairs $(a_n, \mu)$ when $\mu$ is of compact
support, so for such pairs this theorem is vacuous.  However 
since the numbers $E(\mu)$ (respectively  $e(\mu)$) depend only on the operators 
$H_0$ (respectively $H_{0+}$) and the measure $\mu$ we can still
choose sequences $a_n$ and $\mu$ of large support such that the
theorem is non-trivial for such cases.  Of course for $\mu$ of
infinite support, the theorem says that there is always a region 
where pure point spectrum is present.

2. Since we allow for potentials with $a_n$ not vanishing at $\infty$
in all directions, we could not make use of the technique of
Aizenman-Molchanov \cite{am}, for exhibiting pure point spectrum.

3. When $\mu$ has compact support, comparing the smallness of a moment
near the edges of support one exhibits pure point spectrum there by
using the lemma (\ref{lem2.2}) proved by Aizenman
\cite{ma}, comparing the decay rate in energy of the sums of low
powers of the integral kernels of the free operators with some uniform
bounds of low moments of the measure $\mu$ weighted with singular but
integrable factors occurring to the same power.

As in Kirsch-Krishna-Obermeit \cite{kko}, Jaksic-Last \cite{jl}
we also have examples of cases when there is pure a.c. spectrum in an interval 
and pure point spectrum outside.  The part about a.c. spectrum follows as a 
corollary of theorem (\ref{thm1.1}), while the pure point part is
proven as in \cite{kko} (following the proof of their theorem 2.3,
where $\Delta$ can be replaced by any bounded self adjoint operator 
on $\ell^2(\ZZ^d)$ and work through the details, as is done in 
Krishna-Obermeit \cite{ko} Lemma 2.1).  Further when $H_0 = \Delta$,
the Jaksic-Last condition on the mutual non-orthogonality of the
subspaces $\hh_{\omega, n}$, $\hh_{\omega, m}$ is valid whenever
$a_n \neq 0, ~~ n \in \Znu$.

\begin{cor}
\label{cor1.1}
Let $a_n$ be a sequence as in Hypothesis (\ref{hyp1.3}) and $\mu$ as
in Hypothesis (\ref{hyp1.2}).  Let $H_0 = \Delta$. 
Assume further that $a_n \neq 0, ~ n \in \Znu$ goes to zero at $\infty$ 
and $a-supp(\mu) = \RR$.  
Then we have, for almost all $\omega$,
\begin{enumerate}
\item $\sigma_{ac}(H^\omega) = [-2\nu, 2\nu]$.
\item $\sigma_{pp}(H^\omega) = \RR \setminus (-2\nu, 2\nu)$.
\item $\sigma_{sc}(H^\omega) = \emptyset$.
\end{enumerate}
\end{cor}

By explicitly putting in the condition of Jaksic-Last on the
non-orthogonality in the above corollary we also have the corollary
below.  The h given in (1) of the section on examples later satisfy 
the conditions of the corollary.

\begin{cor}
\label{cor1.2}
Let $a_n$ be a sequence as in Hypothesis (\ref{hyp1.3}) and $\mu$ as
in Hypothesis (\ref{hyp1.2}).  Let $H_0$ be a bounded self adjoint
operator coming from h satisfying the hypothesis (\ref{hyp1.1}).
Assume that $a_n \neq 0, ~ n \in \Znu$ goes to zero at $\infty$ 
and $a-supp(\mu) = \RR$.  Suppose further that $\hh_{\omega, n}$ and
$\hh_{\omega, m}$ are not mutually orthogonal for m, n in $\Znu$
almost all $\omega$.  Then we have, for almost all $\omega$,
\begin{enumerate}
\item $\sigma_{ac}(H^\omega) = [-E_-, E_+]$.
\item $\sigma_{pp}(H^\omega) = \RR \setminus (-E_-, E_+)$.
\item $\sigma_{sc}(H^\omega) = \emptyset$.
\end{enumerate}
\end{cor}

\section{Proofs}

In this section we present the proofs of the theorems stated in the 
previous section. 

{\noindent \bf Proof of Theorem (\ref{thm1.0}):}

The statement about the spectrum of $H_0$ follows from the Hypothesis
(\ref{hyp1.1})(1) on the function h.  Each of the functions $h_i$
is continuous on $[0, 2\pi]$, hence has compact range and by the
intermediate value theorem its range is also an interval 
$I_i$.  Since the spectrum of $H_0$ is the algebraic sum of the intervals 
$I_i$, -- if $H_{0j}$ denotes the operator associated with $h_j$ on 
$\ell^2(\TT)$, then $H_0 = H_{01}\otimes I + I\otimes H_{02}\otimes I
+\cdots + I\otimes H_{0\nu}$ hence this fact -- the statement follows.

As for $H_{0+}$ we note that the operator 
$\Delta_+$ restricted to the subspace $\{f \in \ell^2(\ZZ^+): f(0) =
0\}$ is unitarily equivalent to multiplication by $2\cos(\theta),
\theta \in [0, 2\pi]$.  Therefore its spectrum is absolutely
continuous and is $[-2, 2]$.  Since $\Delta_+$ is a rank two
perturbation of this restriction, the absolutely continuous spectrum
of $\Delta_+$ continues to be $[-2, 2]$.  The operator $\Delta_+$ may
have two eigenvalues in addition to this a.c. spectrum, however since
the spectrum of $H_0$ is purely absolutely continuous, the spectrum of
$H_{0+}$ is also purely a.c. and contains 
$\sigma(\Delta_+) + [E_-, E_+]$, with $E_\pm$ as above.
Hence the theorem follows.

{\noindent \bf Proof of Theorem (\ref{thm1.1})}
We prove the theorem for the case $H^\omega$
the proof for the case $H^\omega_+$ 
proceeds along essentially the same lines and we give a sketch of the
proof for that case.  We consider any 
$\lambda \in a-supp (\mu)$, which means that we have 
$$
\sum_{n \in k\Znuplus} \mu(a_n^{-1}(\lambda - \epsilon, \lambda +
\epsilon)) = \infty, ~~ \forall ~~  k \in \ZZ^+, k \neq 0, ~~
\text{and all} ~~\epsilon >0. 
$$
We consider the distance function $|n| = \text{max} |n_i|, i = 1,\cdot
\nu$ on $\Znu$. We consider the events, with $\epsilon > 0$, $m \in
k\Znu$, 
$$
A_{k, m, \epsilon} = \{ \omega : a_m q^\omega(m) \in (\lambda - \epsilon,
~\lambda + \epsilon), ~~ |a_n~ q^\omega(n)| < \epsilon, ~ \forall 0 < |n -
m| < k - 1\}
$$
and 
$$
B_{k, m, \epsilon} = \{ \omega : ~~ |a_n~ q^\omega(n)| < \epsilon, 
~ \forall 0 \leq  |n - m| < k - 1\},
$$
where the index n in the definition of the above sets varies $\Znu$.
Then each of the events $A_{k, m, \epsilon}$ are mutually independent
for fixed k and $\epsilon$ as m varies in $k\Znu$, since the random
variable defining them live in disjoint regions in $\Znu$.   Similarly 
$B_{k, m, \epsilon}$ is a  collection of mutually independent events
for fixed k and $\epsilon$ as m varies in $k\Znu$.  Further these
events have a positive probability of occurance, the probability
bigger than 
$$
\text{Prob}(A_{k, m, \epsilon}) \geq \mu(a_m^{-1}
(\lambda - \epsilon, \lambda + \epsilon))
(\mu(-c ~\epsilon, c~\epsilon))^{(k-1)^{\nu +1}}
$$
and
$$
\text{Prob}(B_{k, m, \epsilon}) \geq (\mu(-c~\epsilon, c~\epsilon))
^{(k-1)^{\nu +1}},
$$
where we have taken $c = \inf_{n \in \Znu} a_n^{-1} > 0$.
The definition of c implies that 
$$
 (-c~\epsilon , c~\epsilon) \subset a_m^{-1}(-\epsilon, \epsilon), ~ 
\forall ~~ m \in \Znu.
$$
Therefore the assumption that $\lambda \in a-supp(\mu)$ implies that
$\forall k \in \ZZ^+\setminus \{0\}$, 
$$
\sum_{m \in k \Znu} \text{Prob}(A_{k, m, \epsilon}) 
\geq 
(\mu(-c~\epsilon, c~\epsilon))^{(k-1)^{\nu +1}}
\sum_{m \in k \Znu} \mu(a_m^{-1}(\lambda - \epsilon, \lambda
+\epsilon)) = \infty 
$$
and similarly
$$
\sum_{m \in k \Znu} \text{Prob}(B_{k, m, \epsilon}) = \infty, ~ 
\forall k \in \ZZ^+\setminus \{0\}. 
$$
Then Borel-Cantelli lemma implies that for all $\epsilon > 0$, 
(setting $R_\epsilon = (\lambda - \epsilon, \lambda+\epsilon)$ and $S_\epsilon
= (-\epsilon, \epsilon)$ and $\Lambda_k(m) = \{n \in \Znu : 0 \leq |n-m|
< k-1\}$), the events 
\begin{equation*}
\Omega(\epsilon, k) = \bigcap_{\substack{m \in I \subset \Znu\\ \#I = \infty}}
\{ \omega : a_m q^\omega(m) \in R_\epsilon,~ 
a_n~ q^\omega(n)\in S_\epsilon, 
~\forall n \in \Lambda_k(m)\setminus \{m\}\} 
\end{equation*}
have full measure.  Therefore the events
$$
\Omega_{1} = \bigcap_{l, k \in \ZZ^+ \setminus \{0\}} \Omega(\frac{1}{l},
k)
$$
has full measure, being countable intersection of sets of full
measure.  Similarly the set
$$
\Omega_2(\epsilon, k) = \bigcap_{\substack{m \in I \subset \Znu\\ \#I = \infty}}
\{ \omega :  a_n~q^\omega(n) \in S_\epsilon, ~\forall n 
\in \Lambda_k(m) \} 
$$
have full measure.  Therefore the events
$$
\Omega_{2} = \cap_{l, k \in \ZZ^+ \setminus \{0\}} \Omega_2(\frac{1}{l},
k)
$$
We take 
$$
\Omega_0 = \Omega_1 \cap \Omega_2
$$
and note that it has full measure.  We use this set for further
analysis.  We denote $H(\lambda) = H_{0} + \lambda P_0$.  Then
suppose $E \in \sigma(H(\lambda))$, then there is a Weyl sequence
$\psi_l$ of compact support, $\psi_l \in \ell^2(\Znu)$ 
such that $\|\psi_l\| = 1$ and
$$
\|(H(\lambda) - E)\psi_l\| < \frac{1}{l}.
$$ 
Suppose the support of $\psi_l$ is contained in a cube of side r(l),
centered at 0. Denote by $\Lambda_k(x)$ a cube of side k centered at x in $\Znu$.
We denote  $V^\omega(n) = a_n q^\omega(n)$, for ease of
writing. We then find cubes $\Lambda_{r(l)}(\alpha_l)$ centered at the points 
$\alpha_l$ such that 
$$
|V^\omega(\alpha_n) - \lambda| < \frac{1}{l}, ~~ |V^\omega(x)| <
\frac{1}{l}, ~~ \forall x \in \Lambda_{r(l)}(\alpha_l) \setminus \{\alpha_l\}.
$$
Now consider $\phi_l(x) = \psi_l(x - \alpha_l)$.  Then by the
translation invariance of $H_0$ we have for any $\omega \in
\Omega_0$,
\begin{equation}
\label{eqn2.00}
\begin{split}
\|(H^\omega - E) \phi_l\| &\leq \|(H_0 + V^\omega(\cdot + \alpha_l)) -
E)\psi_l\|  \\
&\leq \|(H_0 + \lambda P_0 - E) \psi_l\| + \|V^\omega(\cdot+\alpha_l)
- \lambda P_0)\phi_l\|\\ 
&\leq \frac{1}{l} + \frac{1}{l}
\end{split}
\end{equation}
Clearly since $\phi_l$ is just a translate of $\psi_l$, $\|\phi_l\|=1$
for each l. We now have to show that the sequence $\phi_l$ goes to zero weakly.
This is ensured by taking successively $\alpha_k$ large so that
$$
\cup_{j=1}^{k-1} supp (\phi_j) \cap \Lambda_{r(k)}(\alpha_k) =
\emptyset, ~~ \text{and} ~~ supp(\phi_k) \subset \Lambda_{r(k)}(\alpha_k).
$$
This is always possible for each $\omega$ in $\Omega_0$ by
its definition, thus showing that the point E is in the spectrum of 
$H^\omega$, concluding the proof of the theorem.

\vspace{5mm}

{\noindent \bf Proof of Theorem (\ref{thm1.2}:)} 
We first consider the part (1) of the theorem and address the proof of
(2) later. We consider the set
\begin{equation}
\label{defdd}
\dd = \{ \phi \in \ell^2(\Znu) : \text{supp} (\widehat{\phi}) \subset
\TT^\nu \setminus \cc \}, 
\end{equation}
where we denote by $\widehat{\phi}$ the function in $\ell^2(\TT^\nu)$
obtained by taking the Fourier series of $\phi$.
Since the set
$\cc$ is of measure zero, such functions form a dense subset of
$\ell^2(\Znu)$.  We also note that the set $\cc$ is closed in
$\TT^\nu$, thus its complement is open (in fact it is a finite union
of open rectangles ) and each $\phi$ in $\dd$ has compact support
in $\TT^\nu \setminus \cc$.  

We first consider the case when $\mu$ has compact
support.  The general case is addressed at the end of the proof.

Then if we show that the sequence $\Omega(t, \omega)= e^{it H^\omega}
e^{-itH_0}$ is strongly Cauchy for any $\omega$, then standard scattering theory
implies that $\sigma_{ac}(H^\omega) \supset \sigma_{ac}(H_0)$ for that
$\omega$.  We will show below this Cauchy property for a set $\omega$
of full measure.  

To this end we consider the quantity
\begin{equation}
\label{eqn2.1}
\EE \{ \|(\Omega(t, \omega) - \Omega(w, \omega)) \phi \| \}, ~~ \phi \in
\dd 
\end{equation}
and show that this quantity goes to zero as t and w go to $+\infty$.
Then the integrand being uniformly bounded by an integrable function
$\|\phi\|$ and since $\phi$ comes from a dense set, Lebesgue dominated
convergence theorem implies that $\Omega(t, \omega)$ is strongly
Cauchy for every $\omega$ in a set of full measure $\Omega(f)$ that
depends on f in $\ell^2(\Znu)$.  Since $\ell^2(\Znu)$ is separable, we
take the countable dense set $\dd_1$ and consider 
$$
\Omega_0 = \cap_{f \in \dd_1} \Omega(f)
$$  
which also has full measure being a countable intersection of sets of
full measure.   For each $\omega \in \Omega_0$, $\Omega(t, \omega)$
is a family of isometries such that $\Omega(t, \omega)f$ is a strongly
Cauchy sequence for each $f \in \dd_1$, therefore this property also
extends by density of $\dd_1$ to all of $\ell^2(\Znu)$ pointwise in 
$\Omega_0$.   Thus it is enough to show that the quantity in equation
(\ref{eqn2.1}) goes to zero as t and r go to $+\infty$.  

We have the following inequality coming out of Cauchy-Schwarz and
Fubini, for an arbitrary but fixed $\phi \in \dd$.  In the inequality
below we denote, for convenience the operator of multiplication by the
sequence $a_n$ as A and in the first step we write the left hand side
as the integral of the derivative to obtain the right hand side.   

\begin{equation}
\label{eqn2.2}
\begin{split}
\EE \{ \|\Omega(t, \omega)\phi - \Omega(r, \omega) \phi \|\} &\leq 
\EE \{ \|\int_r^t ds ~~ e^{isH^\omega}V^\omega e^{-isH_0} \phi \|\} \\ 
 &\leq \int_r^t ds ~~ \EE \{ \|V^\omega e^{-isH_0} \phi \|\} \\ 
 &\leq \int_r^t ds ~~  \|\sigma A  e^{-isH_0} \phi \|. 
\end{split}
\end{equation}
The required statement on the limit follows if we now shot that the
quantity in the integrand of the last line is integrable in s.
To do this we define the number
\begin{equation}
\label{minvel}
v_{\phi} = \inf_{j} \inf \{ |h_j^\prime(\theta_j)| : \vartheta \in
\text{supp} \widehat{\phi} \}, ~~ \vartheta = (\theta_1, \cdots,
\theta_\nu). 
\end{equation}
We note that since the support of $\widehat{\phi}$ is compact in
$\TT^\nu \setminus \cc$, ${h_j}^\prime, j=1,
\cdots, \nu$ (which are continuous by assumption), have 
non-zero infima there, so $v_\phi$ is strictly positive.
Then consider the inequalities
\begin{equation}
\label{eqn2.3}
\begin{split}
 \|\sigma A  e^{-isH_0} \phi \| &\leq  
\|\sigma A F(|n_j| > v_{\phi} ~ s /4 ~~ \forall j ~~) e^{-isH_0} \phi \|\\   
&~~~~+ \|\sigma A F(|n_j| \leq v_{\phi} ~ s /4 ~~ \text{for some } ~~j) 
e^{-isH_0} \phi \| \\
&\leq \sigma |g(s)| \|\phi\| +    
\sigma \|A\| \|F(|n_j| \leq v_{\phi} ~ s /4, ~~ \text{for some} ~~ j) 
e^{-isH_0} \phi \|, 
\end{split}
\end{equation}
where we have used the notation that F(S) denotes the orthogonal
projection (in $\ell^2(\Znu)$) given by the indicator function of the
set S and used the function g as in the Hypothesis (\ref{hyp1.3})(2) 
which is integrable in s, so the first term is integrable in s.
We concentrate on the remaining term.  
\begin{equation}
\label{eqn2.4}
\|F(|n_j| \leq v_{\phi} ~ s /4, \text{for some} j) 
e^{-isH_0} \phi \| 
\end{equation}
To estimate the term we go to
the spectral representation of $H_0$ and do the computation there as
follows.  
Since $|n_j| \leq v_\phi ~ s /4$ for some j, we may without loss of
generality assume that the index j =1, and proceed with the
calculation.  Let us denote the set $S_1(s) = \{n : 
n_1 \leq v_\phi s/4, ~~ n_j \in \ZZ, ~~ j \neq 1\}$.  In the steps
below we pass to $L^2(\TT^\nu)$ via the Fourier series, (where the normalized 
measure on $\TT^\nu$ is denoted by $d\sigma(\vartheta)$).
\begin{equation}
\label{eqn2.5}
\begin{split}
T &= \|F(|n_1| \leq v_{\phi} ~ s /4) e^{-isH_0} \phi \|\\ 
& = \{\sum_{ n \in S_1(s)} |\langle \delta_n, e^{-isH_0}\phi\rangle|^2\}^{1/2}\\
& = \{\sum_{ n \in S_1(s)} |\int_{\TT^\nu} d\vartheta ~~ e^{-i n\cdot\vartheta -i
s \sum_{j=1}^\nu h_j(\theta_j)}\widehat{\phi}(\vartheta)|^2\}^{1/2}\\
& = \{\sum_{ n \in S_1(s)} |\int_{\TT^\nu} d\vartheta ~~ e^{-i n\cdot\vartheta -i
s \sum_{j=1}^\nu h_j(\theta_j)}\widehat{\phi}(\vartheta)|^2\}^{1/2}\\
& = \{\sum_{n\in\ZZ^{\nu-1}}\sum_{n_1 \leq \frac{v_\phi s}{4}}  
|\int_{\TT^{\nu-1}} \prod_{j=2}^\nu d\sigma(\theta_j) ~~ 
e^{-i\sum_{j=2}^\nu n_j\theta_j + s h_j(\theta_j)} \int_\TT d\sigma(\theta_1)\\
& ~~~
e^{-i(n_1\theta_1 + s h_1(\theta_1))} \widehat{\phi}(\vartheta)
d\sigma(\theta_1) |^2\}^{1/2}.
\end{split}
\end{equation}
We define the function $J(\theta, s, n_1) =  n_1 \theta + s h_1(\theta)$. 
When $\vartheta$ is in the support of $\widehat{\phi}$, 
we have that $|h_1^{\prime}(\theta_1)| \geq v_\phi$, 
by equation (\ref{minvel}).  This 
in turn implies that when $\vartheta = (\theta_1, \cdots, \theta_\nu)
\in \text{supp}\widehat{\phi}$,  
$$
|\frac{\partial}{\partial \theta}J(\theta_1, s, n_1)| 
= | n_1 + s ~ h_1^\prime(\theta_1)| \geq 3 v_\phi ~ s/4  
 ~~ \text{when} ~~  n_1 \leq v_\phi ~ s/4.
$$
We use this fact and do integration by parts twice with respect to the variable 
$\theta_1$ to obtain 
\begin{equation}
\label{eqn2.6}
\begin{split}
T & = \{\sum_{n_1 \leq \frac{v_\phi  s}{4}} \sum_{n \in \ZZ^{\nu-1}} 
|\int_{\TT^{\nu-1}} \prod_{j=2}^\nu d\sigma(\theta_j) ~~ 
e^{-i\sum_{j=2}^\nu n_j\theta_j + s h_j(\theta_j)} \int_\TT d\sigma(\theta_1)\\
& ~~~
e^{-i(n_1\theta_1 + s h_1(\theta_1)} 
\left\{\left(\frac{\partial}{\partial\theta_1}\frac{1}{J^\prime(\theta_1, n_1, s)}
\right)^2 \widehat{\phi}(\vartheta)\right\}
d\sigma(\theta_1) |^2\}^{1/2}.
\end{split}
\end{equation}
We note that the quantity
\begin{equation}
\label{eqn2.08}
\begin{split}
I_1 & = \left(\frac{\partial}{\partial\theta_1}\frac{1}{J^\prime(\theta_1, n_1, s)}
\right)^2 \widehat{\phi}(\vartheta) \\
&= (\frac{-J^{(3)}}{(J^\prime)^3} + \frac{3 J^{(2)}(J^\prime)^2}{(J^\prime)^6})
\widehat{\phi} + \frac{1}{(J^\prime)^2}\frac{\partial^2}{\partial\theta_1^2}
\widehat{\phi} 
+  \frac{-J^{(2)}}{(J^\prime)^3}\frac{\partial}{\partial\theta_1}\widehat{\phi}  
\end{split}
\end{equation}
is in $L^2(\TT^\nu)$.

The assumptions on the lower bound on $J^\prime$ (when $|n_1| \leq
v_\phi s/4$) and the boundedness
of its higher derivatives by $C s$ (which is straightforward to verify
by the assumption on $h_j$) together now yield the bound
$$
T \leq 
\frac{C}{s^2} \left\{ \|\phi\| +
\|\widehat{\frac{\partial}{\partial\theta_1}\phi}\|_{L^2(\TT^\nu)} 
+ \|\widehat{\frac{\partial^2}{\partial\theta_1^2}\phi}\|_{L^2(\TT^\nu)}\right\}. 
$$
Which gives the required integrability.

We proved the case (1) of the theorem assuming that $\mu$ has compact
support.  The case when $\mu$ has infinite support requires only a
comment on the function $e^{-isH_0}\phi$ being in the
domain on $V^\omega$ almost everywhere, when s is finite and for
fixed $\phi \in \dd$.  Once this is ensured the remaining calculations
are the same.  To see the stated domain condition we first note that
for each fixed s, the sequence $(e^{-is H_0}\phi)(n)$ decays faster than any
polynomial, (in $|n|$).  The reason being that, by assumption, $\widehat{\phi}$ is
smooth and of compact support in $\TT^\nu$,  
$|\phi(n)| \leq |n|^{-N}$ for any $N > 0$, as $|n| \rightarrow
\infty$.  On the other hand for $|n-m| > s \|H_0\|$, we have
$$
|e^{-isH_0}(n, m)| \leq \frac{1}{|n-m|^N}, ~~\text{for any } N > 0.
$$ 
These two estimates together imply that 
\begin{equation}
\label{domcon}
\|(1 + |m|)^{2\nu + 2} e^{-isH_0}\phi\| < \infty, ~~\forall ~~ \phi
\in \dd.
\end{equation}

We now consider the events
$$
A_n = \{ \omega : ~ |q^\omega(n)| > |n|^{2\nu + 1} \}
$$
satisfy the condition
$$
\sum_{n \in \Znu} Prob (A_n) < \infty, 
$$
by a simple application of Cauchy-Schwarz and the finiteness of the
second moment of $\mu$.  Hence, by an application of Borel-Cantelli
lemma,  only finitely many events $A_n$ occur
with full measure.  Therefore on  a set of full measure all but
finitely many $q^\omega(n)$ satisfy, $|q^\omega(n)| \leq
|n|^{2\nu+1}$.  Let the set of full measure be denoted by $\Omega_1$.
Then for each $\omega \in \Omega_1$ we have a finite set $S(\omega)$
such that 
$e^{-is H_0}\phi$ is in the domain of the operator
$V^\omega_1 = V^\omega (I - P_{S(\omega)})$, where $P_{S(\omega)}$  
is the orthogonal projection onto the subspace $\ell^2(S(\omega))$,
in view of the equation (\ref{domcon}).  
Then the proof that the a.c. spectrum of the operator 
$$
H_1^\omega = H_0 + V^\omega_1, \forall \omega \in \Omega_1 \cap
\Omega_0 
$$
goes through as before.  Since for each $\omega \in \Omega_1 \cap
\Omega_0$, $H^\omega_1$ differs from $H^\omega$ by a finite rank
operator, its absolutely continuous spectrum is unaffected (by trace
class theory of scattering) and the theorem is proved. 

The statement on the singular part of the spectrum of 
$H^\omega$, is a direct corollary of the theorem \ref{thma.1}. 
We note firstly that since $\delta_n , n \in \Znu$ is an orthonormal basis for
$\ell^2(\Znu)$ it is automatically a cyclic family for $H^\omega$
for every $\omega$.  

Secondly, by assumption, the subspaces $\hh_{\omega, n}$ and
$\hh_{\omega, m}$ are not mutually orthogonal, so the conditions of 
theorem (\ref{thma.1}) are satisfied.  Therefore, since 
the a.c. spectrum of $H^\omega$ contains the
interval $(E_-, E_+)$ almost every $\omega$ the result follows.

(2) We now turn to the proof of part (2) of the theorem.  The
essential case to consider again as in (1) is when  
$\mu$ has compact support, the general case goes through as before.
The proof is again similar to the one in (1), but we need to choose
a dense set $\dd_1$ in the place of $\dd$ properly.

The operator $\Delta_+$ is self adjoint on $\ell^2(\ZZ^+)$ and its
restriction $\Delta_{+1}$ to $\ell^2(\ZZ^+ \setminus \{0\})$ is
unitarily equivalent to multiplication by $2\cos(\theta)$ acting on
the image of $\ell^2(\ZZ^+ \setminus \{0\})$ under the Fourier series
map.  We now consider the operator
$$
H_{0+1} = \Delta_{+1} + H_0
$$
in the place of $H_{0+}$ and show the existence of the Wave operators
$$
W_+ = \text{Slim}_{t \rightarrow \infty}
e^{itH^\omega_+}e^{-itH_{0+1}} 
$$
almost every $\omega$.

We take the set $\dd$ as in equation (\ref{defdd}), $\dd_2$ as in
lemma (\ref{lem2.11}) and define

\begin{equation}
\label{eqn2.7}
\dd_+ = \left\{ \phi  : ~ \phi = \sum_{i,j ~~\text{finite}} \alpha_{ij} \phi_i 
~ \psi_j,
~~ \psi_j \in \dd, ~~ \phi_i \in \dd_{2}, ~~ \alpha_{ij} \in \CC \right\}.  
\end{equation}

Then $\dd_+$ is dense in 
$$
\hh_0 = \{ f \in \ell^2(\Znuplus): f(0, n) = 0 \}.
$$
We then define the minimal velocities for $\phi \in \dd_+$ with 
$w_{\phi_1}$ defined as in lemma (\ref{lem2.11}) for $\phi_1 \in
\dd_2$.
\begin{equation}
\label{minvel2}
\begin{split}
w_{1,\phi} &= \inf_{k} w_{\phi_k} \\ 
w_{2, \phi} &= \inf_{l} \inf_{j} \inf \{ |h_j^{\prime}(\theta_j)| :
 \vartheta \in \text{supp} \widehat{\psi_l} \}\\
v_{\phi} &= \text{min} \{ w_{1, \phi}, w_{2, \phi}\}.
\end{split} 
\end{equation}

Calculating the limits, as in equation (\ref{eqn2.1})
\begin{equation}
\label{eqn2.8}
\begin{split}
\| (e^{itH^{\omega_+}}e^{-itH_{0+1}} -
e^{irH^{\omega_+}}e^{-irH_{0+1}})\phi\| \\
=  \int_r^t ds ~
\| (e^{isH^{\omega_+}}(V^\omega - P_0\Delta_+ +
-\Delta_+P_0+ P_0\Delta_+P_0)e^{-itH_{0+1}}\phi\|,
\end{split}
\end{equation}
where $P_0$ is the operator $p_0\otimes I$, with $p_0$ being the
orthogonal projection onto the one dimensional subspace spanned by the
vector $\delta_{0}$ in $\ell^2(\ZZ^+)$.  We note that by the definition
of $\Delta_+$, the term $P_0\Delta_+P_0$ is zero. 
The estimates proceed as in the proof of (1), after taking averages over
the randomness and taking  $\phi \in \dd_+$. As in that proof it is
sufficient to show the integrability in s of the functions
$$
\|\sigma A e^{-is H_{0+1}}\phi\|, ~~
\||\delta_1><\delta_0|\otimes I e^{-is H_{0+1}}\phi\|, ~~
\||\delta_0><\delta_1|\otimes I e^{-is H_{0+1}}\phi\|, 
$$
respectively.  By the definition of $\dd_+$, any $\phi$ there is
a finite sum of terms of the form
$\phi_j(\theta_1) \psi_j(\theta_2, \cdots, \theta_{\nu+1})$, 
so it is enough to show the integrability when $\phi$ is just  one
such product, say $\phi = \phi_1 \psi_1$. 
Therefore we show the integrability in s of the functions
$$
\|\sigma A e^{-is H_{0+1}}\phi\|, ~~
\||\delta_1><\delta_0|\otimes I e^{-is H_{0+1}}\phi\|, ~~
\||\delta_0><\delta_1|\otimes I e^{-is H_{0+1}}\phi\|, 
$$
for s large, in which case we have  
$$
F(|n_1| > v_\phi s/4) \delta_i = 0, ~~ i = 0, 1 ~~\text{and} ~~
\|\sigma A F(|n_1| > v_\phi s/4)\| \in L^1(1, \infty),
$$
by the hypothesis (\ref{hyp1.3})(4) on the sequence $a_n$.
Therefore it is enough to show the integrability of the norms
$$
\|F(|n_1| < v_\phi s/4) e^{-is \Delta_{0+1}}\phi_1\|, ~~ \forall
\phi_1 \in \dd_2 
$$ 
whose proof is given in the lemma (\ref{lem2.11}) below.

The statement on the absence of singular part of the spectrum of 
$H^\omega+$ in $(E_--2, E_+ +2)$, is as before a direct corollary of 
the theorem \ref{thma.1}, since the set of vectors $\{\delta_n, n =
(0, m), m\in \Znu\}$ is a cyclic family for $H^\omega_+$, for almost
all $\omega$ and $\hh_{\omega, n}$ and $\hh_{\omega, m}$ are not
mutually orthogonal for almost all $\omega$ when m, n are in 
$\{(0, n): n \in \Znu\}$,  
and the fact that the a.c. spectrum of $H^\omega$ contains the
interval $(-2+E_-, 2+E_+)$ almost every $\omega$. 

The lemma below is as in Jaksic-Last \cite{jl2}(lemma 3.11) and the
enlarging of the space in the proof is necessary since there are no non-trivial 
compactly supported functions in $\hh_{0+}$ (all of them being
boundary values of functions analytic in the disk).

\begin{lemma}
\label{lem2.11}
Consider the operator $\Delta_{+1}$ on $\ell^2(\ZZ^+)$.  Then there is
a set $\dd_{2}$ dense in $\ell^{2}(\ZZ^+)$ and a number $w_\phi$ such that 
for $s \geq 1$,
$$
\|F(|n| < w_\phi s/4) e^{-is\Delta_{1+}} \phi\| \leq C |s|^{-2}, ~~
\forall \phi \in  \dd_2.
$$
with the constant C independent of s.
\end{lemma}

{\noindent \bf Proof:} We first consider the unitary map
$\ww$ from $\hh_0$ to a subspace $\WW$ of $\{ f \in \ell^2(\ZZ) : f(0) = 0 \}$, 
given by
\begin{equation}
(\ww f)(n) = \begin{cases} &\frac{1}{\sqrt{2}} f(n), ~~ n > 0 \\
 &-\frac{1}{\sqrt{2}} f(-n), ~~ n < 0.
\end{cases}
\end{equation}
Then the range of $\ww$ is a closed subspace of $\ell^2(\ZZ)$ and
consists of functions
$$
\WW = \{ f \in \ell^2(\ZZ) : f(n) = - f(-n) \}.
$$
Under the Fourier series map this subspace goes to 
$$
\widehat{\WW} = \{ \phi \in L^2(\TT): \phi(\theta) = -\phi(-\theta) \}
$$
so that the functions here have mean zero.  Then under the map from
$\ell^2(\ZZ^+\setminus \{0\})$ to $\widehat{\WW}$ obtained by composing 
$\ww$ and the Fourier series
map, the operator $\Delta_{1+}$ goes to multiplication by $2\cos(\theta)$.
We now choose a set
$$
\dd_1 = \{ \phi \in \widehat{\WW} : \text{supp} (\phi) \subset \TT \setminus
\{0, \pi\} \},
$$
and define the number 
$$
w_{\phi} = \inf \{ |2\sin(\theta)| : \theta \in \text{supp}(\phi) \},
$$
for each $\phi \in \dd_1$.  We denote by $\dd_2$ all those functions
whose images under the composition of $\ww$ and the Fourier series
lies in $\dd_1$.  The density of $\dd_2$ in $\ell^2(\ZZ^+\setminus
\{0\})$ is then clear.  We shall simply denote by $f_\phi$ elements in 
$\dd_2$ whose images in $\dd_1$ is $\phi$.  Given a $\phi \in \dd_1$
and a $w_\phi$ we see that   
$$
\|F(|n| \leq w_\phi s/4) e^{-is\Delta_{1+}}f_\phi\|^2  
= \sum_{|n| < w_\phi s/4} |\int_{\TT} d\sigma(\theta) e^{-in\theta -i2s\cos(\theta)}
\phi(\theta)|^2
\leq C |s|^{-4},
$$
by a simple integration by parts, done twice, using the condition that
$||n| + 2 s \sin(\theta)| > w_\phi s /4$ in the support of $\phi$.

{\noindent \bf Proof of theorem (\ref{thm1.3}):}

The proof of this theorem is based on a technique of Aizenman \cite{ma}.
We break up the proof into a few lemmas.  First we show that the free
operators $H_0$ and $H_{0+}$ have resolvent kernels with some
summability properties, for energies in their resolvent set.

\begin{lemma}
\label{lem2.1}
Consider a function h satisfying the Hypothesis (\ref{hyp1.1})(1-3)
and consider the associated operators $H_0$ or $H_{0+}$.  Then
for all $s \geq \nu/(3\nu + 3)$, 
$$
\sup_{n \in \Znu} \sum_{n\in \Znu} 
|<\delta_n, (H_0 - E)^{-1}\delta_m>|^{s} < C(E), 
$$
and $C(E) \rightarrow 0, |E| \rightarrow \infty$.
Similarly we also have for all $s > \nu/(3\nu+3)$,
$$
\sup_{n \in \Znuplus } \sum_{n\in \Znu} 
|<\delta_n, (H_{0s} - E)^{-1}\delta_m>|^{s} < C(E), 
$$
\end{lemma}
{\noindent \bf Proof:}  We will prove the statement for $H_0$ the
proof for $H_{0+}$ is similar.  We write the expression for the
resolvent kernel in the Fourier transformed representation 
(we write the Fourier series of an $\ell^2(\Znu)$ function as
$\widehat{u}(\vartheta) = \sum_{n \in \Znu} e^{in\cdot\vartheta}u(n)$),
use the Hypothesis (\ref{hyp1.1}(1)(3) (which ensures that the
boundary terms are zero), and integrate by
parts $3\nu+3$ times with respect to the variable $\theta_j$
(recall that $\vartheta = (\theta_1, \cdots, \theta_\nu)$), to get the inequalities 
\begin{equation}
\begin{split}
<\delta_n, (H_0 - E)^{-1}\delta_m>
&= \int_{\TT^\nu} d\sigma(\vartheta) ~~ e^{i(m-n)\cdot\vartheta} 
~ (h(\vartheta) - E)^{-1} 
= \frac{(i)^{3\nu+3}}{((m-n)_j)^{3\nu+3}}\\
& ~~~ \times \int_{\TT^\nu} d\sigma(\vartheta) 
~~ e^{i(m-n)\cdot\vartheta} 
~ \frac{\partial^{3\nu+3}}{\partial \theta_j^{3\nu+3}}(h(\vartheta) - E)^{-1}.
\end{split}
\end{equation}
Where we have chosen the index j such that $|(m-n)_j| \geq |m-n|/\nu$
and assumed that $m\neq n$ (when m=n the quantity is just bounded). 
Let us set 
$$
C_0(E) = \text{max}~~\{\sup_{\vartheta \in \TT^\nu}  
|\frac{\partial^{3\nu+3}}{\partial \theta_j^{3\nu+3}}(h(\vartheta) - E)^{-1}|,
~ |(h(\vartheta) - E)^{-1}| \}.
$$
It is easy to see that since the function h is of compact range and all its $3\nu+3$
partial derivatives are bounded, by hypothesis $C(E)$ goes to zero as
$|E|$ goes to $\infty$.
We then get the bound for any $s > \nu /(3\nu+3)$.   
$$
|<\delta_n, (H_0 - E)^{-1}\delta_m>| \leq
\frac{\nu^{3\nu+3}}{|m-n|^{3\nu+3}} C_0(E), 
$$
Given this estimate we have
\begin{equation}
\begin{split}
\sup_{n \in \Znu} \sum_{n\in \Znu} 
|<\delta_n, (H_0 - E)^{-1}\delta_m>|^s 
&\leq C_0(E)^s(\sup_{n \in \Znu} (1 + \sum_{\substack{n\in \Znu\\ m\neq n}} 
|\frac{\nu^{s(3\nu+3)}}{|m-n|^{s(3\nu+3)}}|)) \\ 
&\leq C_0(E)^s(1 + \sum_{n\in \Znu, m\neq 0} 
|\frac{\nu^{s(3\nu+3)}}{|m|^{s(3\nu+3)}}|) \\ 
&\leq C_0(E)^s C(s).  
\end{split}
\end{equation}
where $C(s)$ is finite since $|m|^{s(3\nu+3|)}, ~m\neq 0$ is a summable function
when $s(3\nu+3) > \nu$.

{\noindent \bf Proof of theorem (\ref{thm1.3}):}
We prove the theorem only for the case $H^\omega$ the proof of the
other case is similar.

By the Hypothesis 
(\ref{hyp1.3})(2)on the finiteness of the
second moment of $\mu$ we see that $\int d\mu(x) ~ |x| < \infty$, so
that we can set $\tau = 1$ in the lemma (\ref{lem2.2}).  Since the assumption
in the theorem ensures the boundedness of the density of $\mu$ we can
also set $q = \infty$ in the lemma (\ref{lem2.2}) with then $Q^{1/1+q} =
\|d\mu/dx\|_\infty$.  Then in the lemma (\ref{lem2.2}) the constant C is
given by
$$
C(Q, \frac{\kappa}{1 - 2\kappa}, \infty) = 1 + \frac{2 \kappa Q}{1 - \kappa}.
$$      
The condition on the constant $\kappa$ becomes 
$$
\kappa < 1/3.
$$
Below we choose a s satisfying $\frac{\nu}{3\nu +3} < s < 1/3$, and
consider the expression
$$
G(\omega, z, n, m) = <\delta_n, (H^\omega - z)^{-1} \delta_m>, ~~
G(0, z, n, m) = <\delta_n, (H_0 - z)^{-1} \delta_m>
$$
where we take $z= E + i\epsilon$ with $\epsilon > 0$ to zero
Then by the resolvent equation we have
\begin{equation}
\label{eqn2.19}
G(\omega, z, n, m) = G(0, z, n, m) + \sum_{l \in Znu} G(\omega, z, n,
l) V^\omega(l)G(0, z, l, m)
\end{equation}
We denote by
$$
G_l(\omega, z, n, m) = <\delta_n, (H^\omega -V^\omega(l)P_l - z)^{-1} \delta_m>, ~~
$$
where $P_l$ is the orthogonal projection onto the subspace generated
by $\delta_l$.  Then using the rank one formula
$$
G(\omega, z, n, l) = \frac{\frac{G_l(\omega, z, n, l)}{G_l(\omega, z, l, l)}}
{V^\omega(l) + G_l(\omega, z, l, l)^{-1}}
$$
whose proof is again by resolvent equation, we see that equation
(\ref{eqn2.19}) can be rewritten as 
\begin{equation}
\label{eqn2.20}
\begin{split}
G(\omega, z, n, m) &= G(0, z, n, m)\\ 
&~~ + \sum_{l \in \Znu} 
( \frac{\frac{G_l(\omega, z, n, l)}{G_l(\omega, z, l, l)}}
{V^\omega(l) + G_l(\omega, z, l, l)^{-1}})
 V^\omega(l)G(0, z, l, m)
\end{split}
\end{equation}
Raising both the sides to power s (noting that $s < 1$ so the
inequalities are valid), we get 
\begin{equation}
\label{eqn2.21}
\begin{split}
|G(\omega, z, n, m)|^s &= |G(0, z, n, m)|^s \\ &~~+ \sum_{l \in \Znu} 
|( \frac{\frac{G_l(\omega, z, n, l)}{G_l(\omega, z, l, l)}}
{V^\omega(l) + (G_l(\omega, z, l, l)^{-1}})|^s
 |V^\omega(l)|^s |G(0, z, l, m)|^s
\end{split}
\end{equation}
Now observing that $G_l$ is independent of the random variable
$V^\omega(l)$, we see that 
\begin{equation}
\label{eqn2.22}
\begin{split}
\EE (|G(\omega, z, n, m)|^s) &= |G(0, z, n, m)|^s \\ &~~+ \sum_{l \in \Znu} 
\EE(|( \frac{\frac{G_l(\omega, z, n, l)}{G_l(\omega, z, l, l)}}
{V^\omega(l) + (G_l(\omega, z, l, l)^{-1}})|^s
 |V^\omega(l)|^s) |G(0, z, l, m)|^s
\end{split}
\end{equation}
This then becomes, integrating with respect to the variable
$q^\omega(l)$, remembering that $V^\omega(l) = a_l q^\omega(l)$, 
\begin{equation}
\label{eqn2.23}
\begin{split}
\EE (|G(\omega, z, n, m)|^s) &= |G(0, z, n, m)|^s \\ &~~+ \sum_{l \in \Znu} 
\EE( |\frac{G_l(\omega, z, n, l)}{G_l(\omega, z, l, l)}|^s \\ &~~~
\times ~  \int
(d\mu(x) ~ \frac{|x|^s}{ |x  + a_l^{-1}G_l(\omega, z, l, l)^{-1}|^s})
 |G(0, z, l, m)|^s
\end{split}
\end{equation}
which when estimated using the lemma (\ref{lem2.2}) yields
\begin{equation}
\label{eqn2.24}
\begin{split}
\EE (|G(\omega, z, n, m)|^s) &= |G(0, z, n, m)|^s \\ & ~~+ \sum_{l \in \Znu} 
K_s \EE( |\frac{G(\omega, z, n, l)}{G_l(\omega, z, l, l)}|^s \\ &~~~
\times ~\int
(d\mu(x) ~ \frac{1}{| x  + a_l^{-1}G_l(\omega, z, l, l)^{-1}|^s} )
 |G(0, z, l, m)|^s, 
\end{split}
\end{equation}
where $K_s$ is the constant
appearing in lemma (\ref{lem2.2}) with $\kappa$ set equal to s.
We take $K = (\sup_{n}|a_n|^s)K_s$, and rewrite the above equation to
obtain
\begin{equation}
\label{eqn2.25}
\EE (|G(\omega, z, n, m)|^s) = |G(0, z, n, m)|^s + \sum_{l \in \Znu} 
K \EE(|( G(\omega, z, n, l)|^s |G(0, z, l, m)|^s
\end{equation}
We now sum both the sides over m, set 
$$
I = \sum_{m \in \Znu} \EE (|G(\omega, z, n, m)|^s) 
$$
and obtain the inequality
$$
I \leq 
\sum_{m \in \Znu} |G(0, z, n, m)|^s + \sup_{l \in \Znu} \sum_{m \in \Znu} 
K I |G(0, z, l, m)|^s. 
$$
Therefore when there is an interval (a, b) in which  
\begin{equation}
\label{eqn2.26}
K \sup_{l \in \Znu} \sum_{m \in Znu} |G(0, z, l, m)|^s < 1,  ~~ E
\in (a, b), 
\end{equation}
we obtain that 
$$
\int_a^b dE ~ \sum_{m \in \Znu} \EE (|G(\omega, E+i0, n, m)|^s) < \infty,  
$$
by an application of Fatou`s lemma implying that for almost all $E \in (a, b)$
and almost all $\omega$,  we have the finiteness of
$$
\sum_{m \in \Znu} |G(\omega, E+i0, n, m)|^2 < \infty,  
$$ 
satisfying the Simon-Wolff \cite{sw} criterion. This shows that
(the proof follows as in  Theorems II.5- II.6 Simon \cite{bs})
the measures 
$$
\nu_n^\omega(\cdot) = <\delta_n, E_{H^\omega}(\cdot) \delta_n>
$$
are pure point in (a, b) almost every $\omega$.  This happens for all
n, hence the total spectral measure of $H^\omega$ itself is pure point
in (a, b) for almost all $\omega$.

There are two different ways to fix the critical energy $E(\mu)$ now.
Firstly if K is large, then
In view of the lemma (\ref{lem2.1}) (by which $C_0(E) \rightarrow 0, ~ |E|
\rightarrow \infty$) and the fact that K is finite (by lemma
\ref{lem2.2}) 
\begin{equation}
\label{eqn2.27}
K \sup_{l \in \Znu} \sum_{m \in Znu} |G(0, z, l, m)|^s 
\leq K C_0(E)^s C(s) < 1, ~~  |E| \rightarrow \infty. 
\end{equation}
Therefore there is a
large enough $E(\mu)$ such that for all intervals (a, b) in 
$(-\infty, -E(\mu)) \cup (E(\mu), \infty)$, the condition in equation
(\ref{eqn2.24}) is satisfied. 

On the other hand if the moment $B = \int |x| ~ d\mu(x)$ is very
small, then we can choose $E(\mu)$ by the condition, 
$$
K C_0(E)C_s < 1,
$$
even when $C_0(E) > 1$, since it is finite for E in the resolvent set
of $H_0$ by lemma (\ref{lem2.1}).

\section{Examples}

In this section we present some examples of the operators $H_0$
considered in the theorems.  We only give the functions h stated in
the Hypothesis (\ref{hyp1.1}). 

\begin{itemize}
\item Examples of operators $H_0$.

\begin{enumerate}
\item $h(\vartheta) = \sum_{i=1}^\nu 2 \cos(\theta_i)$, corresponds to
the usual discrete Schr\"odinger operator and it is obvious that the
hypothesis (\ref{hyp1.1}) are satisfied.  The Jaksic-Last condition 
(\ref{thma.1}) on
mutual non-orthogonality of the subspaces generated by $H_0$ and
$\delta_n$ for different n in $\Znu$ are also satisfied, by an
elementary calculation taking powers of $H_0$ depending upon a pair of
vectors $\delta_n$ and $\delta_m$, since the operator $H_0$ is given
by $T + T^{-1}$, with T being the bilateral shift on $\ell^2(\ZZ)$.
\item  $h(\vartheta) = \sum_{i=1}^\nu h_i(\theta_i), h_i(\theta_i) = 
\sum_{k=1}^{N(i)} \cos(k\theta_i), ~~ N(i) < \infty$.  For this operator (1)
and (3) are satisfied since $h_i$ is $C^\infty$ in $\theta_i$ while
the function and all its derivatives satisfy $h^{(j)}(0) = h^{(j)}(2\pi)$,
whose values are 0 for j odd and $(-1)^{j/2} \sum_{k=1}^{N(i)} k^j$ for
j even. Hence (3) of the hypothesis (\ref{hyp1.3}) is satisfied . 
Each of $h_i$ is a trigonometric
polynomial, and its derivative is also a trigonometric polynomial and
hence has only finitely many zeros on the circle.

The condition in Jaksic-Last condition theorem (\ref{thma.1}) 
on mutual non-orthogonality is again elementary to verify in this case.

\item Consider the functions
$$
h_i(\theta_i) = \theta_i^{3\nu +4}(2\pi - \theta_i)^{3\nu +4}
$$
and take $h = \sum_{i=1}^\nu h_i(\theta)$.  Clearly these  are in  
$C^{3\nu+3}$.  And by construction taking $3\nu+3$ derivatives of
these functions always will have a factor $\theta_i (2\pi - \theta_i)$
ensuring that they vanish at the boundary points. 
\end{enumerate}

\item Examples of pairs ($a_n$, $\mu$).

We give next some examples of sequences $a_n$ satisfying the Hypothesis
(\ref{hyp1.2}) such that 
$$
supp(\mu) = a-supp(\mu).
$$
We consider $\nu \geq 2$ and the sequence $a_n = (1+ |n_1|)^{\alpha}, ~ \alpha <
-1$.
Then we have that 
$$
k\Znu \cap \{(0, n) : n \in \ZZ^{\nu-1}\} =  
\{(0, n) : n \in k\ZZ^{\nu-1}\}  
$$
and $a_{(0,n)}^{-1}(a,~ b) = (a,~ b)$ for any
interval (a,~ b) and any $n \in \ZZ^{\nu - 1}$.  Therefore for any 
positive integer k, we have 
$$
\sum_{m \in k\ZZ^\nu} \mu(a_m^{-1}(a,~ b))  \geq  
\sum_{m \in k\ZZ^{\nu-1}} \mu((a,~ b))  = \infty 
$$
whenever $\mu((a,~ b)) >0$.

\item Examples of measures $\mu$ with small moment.

We next give an example of an absolutely continuous measure of
compact support such that the Aizenman condition (in lemma
(\ref{lem2.2}) is satisfied.  We use the notation used in that lemma
for the example.

We consider numbers $0 < \epsilon, \delta < 1$, R and let $\mu$ be given by
\begin{equation}
d\mu(x)/dx = \begin{cases} & \frac{1-\epsilon}{\delta}, ~~ 0 \leq x \leq
\delta,\\
&\frac{\epsilon}{R - \delta}, ~~ \delta < x \leq R, \\
& 0, ~~ \text{otherwise}.
\end{cases}
\end{equation}
Then $\mu$ is an absolutely continuous probability measure and 
$$
Q \leq \frac{1}{\delta} + \frac{1}{R -\delta}.  
$$
We take $\tau = 1$, then the moment B is bounded by, 
$$
B \leq (1 - \epsilon)\delta + (R + \delta)\epsilon/2.
$$
Now if we fix R large and choose $\epsilon = 1/R^3$ and $\delta = 1/R^2$,
we obtain an estimate
$$
B^\kappa \leq \frac{2^\kappa}{R^{2\kappa}} ~~ \text{and} ~~ B^\kappa Q^{1 - 2\kappa} \leq 
8 R^{2 - 6\kappa}.
$$
Taking $\kappa = s$ in the theorem and noting that $s < 1/3$ implies
$2 - 6s < 0$ so that both the terms above go to zero as R goes to 
$\infty$. We see that by taking $\mu$ with large support but small
moment, we can make the constant K in the lemma (\ref{lem2.2}) as small as
we want.  This in particular means that in the theorem (\ref{thm1.3})
given a energy $E_0$ outside the
spectrum of $H_0$ we can find a measure $\mu$ which is absolutely
continuous of small moment such that $K$ is smaller than
$C_0(E_0)^s C_s$ in the proof of theorem (\ref{thm1.3}) and hence
$E(\mu) < |E_0|$. 
We can use such measures to give examples of
operators with compact spectrum with both a.c. spectrum and pure point
spectrum present but in disjoint regions.

\item Example when Jaksic-Last condition is violated.

We finally give examples where Jaksic-Last condition is violated and
yet the conclusion of their theorem is valid.

Consider $\nu = 1$, for simplicity, and let $h(\theta) = 2
\cos(2\theta)$.  Then the associated $H_0$ has purely a.c. spectrum in
[-2, 2] and we see that the operator $H_0 = T^2 + T^{-2}$ if T is the
bilateral shift acting on $\ell^2(\ZZ)$.  Then if we consider the
operators $H^\omega = H_0 + V^\omega$, and the cyclic subspaces
$\hh_{\omega, 1}, \hh_{\omega, 2}$ generated by the $H^\omega$ and the
vectors $\delta_1, \delta_2$ respectively, such an operator satisfies
$$
\hh_{\omega, 1} \subset \ell^2(\{1\}+2\ZZ), ~~ 
\hh_{\omega, 2} \subset \ell^2(\{1+1\}+2\ZZ), ~~ a.e. \omega 
$$
We then have 
$$
\hh_{\omega, 1} \subset \ell^2(\{2n+1, ~ n \in \ZZ\}), ~~ 
\hh_{\omega, 2} \subset \ell^2(2\ZZ), ~~  a.e. \omega.
$$
The subspaces $\ell^2(\{n: ~ n~ odd\})$ and $\ell^2(2\ZZ)$ are generated by the
families $\{\delta_k,~ k ~odd\}$ and $\{\delta_k, ~ k ~even\}$
respectively.  (We could have taken any odd integer k in the place of 1
to do the above) 

These two are invariant subspaces of
$H^\omega$ which are mutually orthogonal, a.e. $\omega$.  Therefore
the Jaksic-Last theorem is not directly valid.  However, by
considering the restrictions of $H^\omega$ to these two subspaces, one
can go through their proof in these subspaces to again obtain the
purity of a.c. spectrum for such operators when they exist. 

We consider two examples to illustrate the point, for which we let
$q^\omega(n)$ denote a collection of i.i.d. random variables with
an absolutely continuous distribution $\mu$ of compact support in $\RR$,
its support containing 0.
\begin{enumerate}
\item If $V^\omega(n) = a_n q^\omega(n)$, with $0< a_n <
(1 + |n|)^{-\alpha}, \alpha > 0$, we see that there is pure a.c. spectrum in 
[-2, 2], a.e. $\omega$ by applying trace class theory of scattering.

\item On the other hand if, with $0 < a_n < (1 + |n|)^{-\alpha}, \alpha
>1$, 
\begin{equation*}
V^\omega(n) = 
\begin{cases}
&a_n q^\omega(n), ~ n ~ \text{odd}\\ 
& q^\omega(n), ~ n ~ \text{even}, 
\end{cases}
\end{equation*}
then there is dense pure point spectrum embedded in the a.c. spectrum
in [-2, 2].
\end{enumerate}
We can give similar, but non trivial, examples in higher dimensions 
but we leave it to the reader.
\end{itemize}

\section{Appendix}

In this appendix we collect two theorems we use in this paper.  One is
a lemma of Aizenman \cite{ma} and another a theorem of Jaksic-Last
\cite{jl}.

The first lemma and its proof are those of Aizenman \cite{ma}(lemma
A.1) which reproduce below (with some modifications in the form we
need), with a slight change in notation (we in particular call the 
number s in Aizenman`s lemma as $\kappa$),

\begin{lemma}[Aizenman]
\label{lem2.2}
Let $\mu$ be an absolutely continuous probability measure whose
density $f$ satisfies $\int_\RR dx |f(x)|^{1+q} = Q < \infty$ 
for some $q > 0$.   Let $0< \tau \leq 1$ and suppose
$B \equiv \int_\RR d\mu(x) ~ |x|^\tau < \infty$.  Then for any
$$
\kappa < \left[ 1 + \frac{2}{\tau} + \frac{1}{q} \right]^{-1}
$$ 
we have
$$
\int_\RR d\mu(x) ~~ \frac{|x|^\kappa}{|x - \alpha|^\kappa} < K_\kappa 
\int_\RR d\mu(x) ~
\frac{1}{|x - \alpha|^\kappa}, ~~ \text{for all } ~~ \alpha \in \RR,
$$
with $K_\kappa$ given by
$$
K_\kappa = B^{\frac{\kappa}{\tau}}(2^{1+2\kappa} + 4)
\left[ B^{1 - \frac{\kappa}{\tau}} +
B^{\frac{\kappa}{\tau}} ~ C(Q, \frac{\kappa}{1-\frac{2\kappa}{\tau}},
q)^{\frac{\tau - 2\kappa}{\tau}}\right] < \infty.
$$
\end{lemma}

{\noindent \bf Remark:} 1. We see from the explicit form of the
constant $K_\kappa$ that the moment B can be made sufficiently small by the
choice of $\mu$ even when its support is large.  This will ensure
that in some models of random operators, the region where the
Simon-Wolff criterion is valid extends to the region in the spectrum.
This is the reason for our writing $K_\kappa$ in this form.  

{\noindent \bf Proof:} The strategy employed in proving the lemma is
to consider the ratio
$$
\frac{\int_\RR d\mu(x) ~~ \frac{|x|^\kappa}{|x - \alpha|^\kappa}}{ \int_\RR d\mu(x) ~
\frac{1}{|x - \alpha|^\kappa}}
$$ 
and obtain upper bounds for the numerator and lower bounds for the
denominator.

Note first that  B finite and $\kappa < \tau$
implies that $|x - \alpha|^\kappa$ is integrable and we have
\begin{equation}
\label{eqn2.10}
\int_a^b f(x) dx \leq Q^{\frac{1}{1+q}}|b-a|^\frac{q}{1+q}
\end{equation}  
by H\"older inequality. 
Hence
\begin{equation}
\label{eqn2.11}
\begin{split}
\int d\mu(x) ~ \frac{1}{|x - \alpha|^\kappa} &\leq 
1 + \int_1^\infty dt ~ \mu(\{x: \frac{1}{|x - \alpha|^\kappa} \geq t\})
\\ 
&\leq 1 + \frac{\kappa (2^q Q)^{\frac{1}{1+q}}}{\frac{q}{1+q} - \kappa}
\\ &\equiv C(Q, \kappa, q),
\end{split}
\end{equation}
where the integral is estimated using the estimate in equation
(\ref{eqn2.10}).  

{\noindent \bf Consider the region $|\alpha| > (2B)^\frac{1}{\tau}$:}

We then estimate for fixed $\alpha$ the
contributions from the regions $|x| \leq |\alpha|/2$ and $|x| >
|\alpha|/2$ to obtain
\begin{equation}
\label{eqn2.12}
\begin{split}
\int d\mu(x) ~ \frac{|x|^\kappa}{|x-\alpha|^\kappa} 
&\leq \frac{2^\kappa}{|\alpha|^\kappa} ( \int d\mu(x) ~ |x|^\kappa + 
\int d\mu(x) ~ \frac{|x|^{2\kappa}}{|x-\alpha|^\kappa})\\
&\leq \frac{2^\kappa}{|\alpha|^\kappa} ( B + B^{\frac{2\kappa}{\tau}}~C(Q,
\frac{\kappa}{1- 2\kappa/\tau}, q)^{\frac{\tau - 2\kappa}{\tau}}),
\end{split}
\end{equation}
with $\kappa$ chosen so that $\kappa/(1-2\kappa/\tau) < q/(1 + q)$.
(Here we have explicitly calculated the p occurring in the lemma of Aizenman
in terms of $\kappa$ and $\tau$).  For  a fixed $\tau$ and $q$ this
condition is satisfied whenever $\kappa$ satisfies the inequality 
stated in the lemma.  

The lower bounds on $\int d\mu(x) 1/|x - \alpha|^\kappa$ is obtained
first by noting that $B < \infty$ implies 
$$
\mu(\{x : |x|^\tau > (2B)\}) \leq \frac{1}{2}.
$$
Since $|\alpha|  > (2B)^\frac{1}{\tau}$, we have the trivial estimate
\begin{equation}
\label{eqn2.13}
\begin{split}
\int d\mu(x) \frac{1}{|x - \alpha|^\kappa} & \geq 
\int_{|x|>(2B)^{\frac{1}{\tau}}} d\mu(x) \frac{1}{|x - \alpha|^\kappa} + 
\int_{|x|\leq (2B)^{\frac{1}{\tau}}} d\mu(x) \frac{1}{|x - \alpha|^\kappa} \\ 
&\geq \int_{|x|\leq (2B)^{\frac{1}{\tau}}} d\mu(x) \frac{1}{|x - \alpha|^\kappa} \\ 
&\geq \frac{1}{2(|\alpha| + (2B)^\frac{1}{\tau})}.
\end{split}
\end{equation}
Putting the inequalities in (\ref{eqn2.12}) and (\ref{eqn2.13}) together
we obtain, (remembering that $|\alpha| > (2B)^{\frac{1}{\tau}}$),
\begin{equation}
\label{eqn2.14}
\frac{\int_\RR d\mu(x) ~~ \frac{|x|^\kappa}{|x - \alpha|^\kappa}}{ \int_\RR d\mu(x) ~
\frac{1}{|x - \alpha|^\kappa}} \leq 2^{1 + 2\kappa}
B^{\frac{\kappa}{\tau}} \left[ B^{1 - \frac{\kappa}{\tau}} +
B^{\frac{\kappa}{\tau}} ~ C(Q, \frac{\kappa}{1-\frac{2\kappa}{\tau}},
q)^{\frac{\tau - 2\kappa}{\tau}}\right]. 
\end{equation}

{\noindent \bf We now consider the region $|\alpha| < (2B)^{\frac{1}{\tau}}$ :}

Estimating as in equation (\ref{eqn2.12}) but now splitting the
region as $|x| \leq (2B)^{\frac{1}{\tau}}$ and $|x| >
(2B)^{\frac{1}{\tau}}$, we obtain the analogue of the estimate
in equation (\ref{eqn2.12}), in this region of $\alpha$ as
\begin{equation}
\label{eqn2.15}
\begin{split}
\int d\mu(x) ~ \frac{|x|^\kappa}{|x-\alpha|^\kappa} 
&\leq \frac{1}{(2B)^{\frac{1}{\tau}}} ( \int d\mu(x) ~ |x|^\kappa + 
\int d\mu(x) ~ \frac{|x|^{2\kappa}}{|x-\alpha|^\kappa})\\
&\leq \frac{1}{(2B)^{\frac{1}{\tau}}} ( B + B^{\frac{2\kappa}{\tau}}~C(Q,
\frac{\kappa}{1- 2\kappa/\tau}, q)^{\frac{\tau - 2\kappa}{\tau}}),
\end{split}
\end{equation}
Similarly the estimate for the denominator term is done as in
equation (\ref{eqn2.13}), 
\begin{equation}
\label{eqn2.16}
\begin{split}
\int d\mu(x) \frac{1}{|x - \alpha|^\kappa} & \geq 
\int_{|x|>(2B)^{\frac{1}{\tau}}} d\mu(x) \frac{1}{|x - \alpha|^\kappa} + 
\int_{|x|\leq (2B)^{\frac{1}{\tau}}} d\mu(x) \frac{1}{|x - \alpha|^\kappa} \\ 
&\geq \int_{|x|\leq (2B)^{\frac{1}{\tau}}} d\mu(x) \frac{1}{|x - \alpha|^\kappa} \\ 
&\geq \frac{1}{2((2B)^\frac{1}{\tau} + (2B)^\frac{1}{\tau})} \\
& = \frac{1}{4(2B)^\frac{1}{\tau}}.
\end{split}
\end{equation}
Using the above two inequalities we obtain the estimate,
\begin{equation}
\label{eqn2.17}
\frac{\int_\RR d\mu(x) ~~ 
\frac{|x|^\kappa}{|x - \alpha|^\kappa}}{ \int_\RR d\mu(x) ~
\frac{1}{|x - \alpha|^\kappa}} \leq 
4\left[ B^{1 - \frac{\kappa}{\tau}} +
B^{\frac{\kappa}{\tau}} ~ C(Q, \frac{\kappa}{1-\frac{2\kappa}{\tau}},
q)^{\frac{\tau - 2\kappa}{\tau}}\right], 
\end{equation}
when $|\alpha| \leq (2B)^{\frac{1}{\tau}}$.  
Using the  inequalities (\ref{eqn2.14}) and (\ref{eqn2.16})
obtained for these two regions of values of $\alpha$ we finally get
\begin{equation}
\frac{\int_\RR d\mu(x) ~~ \frac{|x|^\kappa}{|x - \alpha|^\kappa}}{ \int_\RR d\mu(x) ~
\frac{1}{|x - \alpha|^\kappa}} \leq 
B^{\frac{\kappa}{\tau}}(2^{1+2\kappa} + 4)\left[ B^{1 - \frac{\kappa}{\tau}} +
B^{\frac{\kappa}{\tau}} ~ C(Q, \frac{\kappa}{1-\frac{2\kappa}{\tau}},
q)^{\frac{\tau - 2\kappa}{\tau}}\right],
\end{equation}
for any $\alpha \in \RR$.

We next state a theorem (Corollary 1.1.3) of Jaksic-Last \cite{jl} without proof, 
its proof is as in Corollary 1.1.3 of Jaksic-Last \cite{jl}.  We state it
in the form we use in this paper.

\begin{thm}
\label{thma.1}[Jaksic-Last]
Suppose $\hh$ is a separable Hilbert space and A a bounded self
adjoint operator.  Suppose $\{\phi_n\}$ are  
normalized vectors  and let $P_n$ denote the orthogonal projection on
to the one dimensional subspace generated by each $\phi_n$.  
Let $q^\omega(n)$ be independent random variables with
absolutely continuous distributions $\mu_n$.  Consider 
$$
A^\omega = A + \sum_n q^\omega(n) P_n, ~~ a.e. ~~ \omega
$$
Suppose that the following conditions are valid  
\begin{enumerate}
\item The  family $\{\phi_n\}$ is a cyclic family for $A^\omega$ a.e.
$\omega$.
\item Let $\hh_{\omega, n}$ denote the cyclic subspace generated by $A^\omega$
and $\phi_n$.  Then the cyclic subspaces $\hh_{\omega, n}$ and $\hh_{\omega,m}$,
are not orthogonal,
\end{enumerate}
Then whenever there is an
interval $(a, ~b)$ in the absolutely continuous spectrum of $A^\omega
= A + \sum_{n} q^\omega(n) P_n$, almost all $\omega$, we have
$$
\sigma_{s}(A^\omega) \cap (a, b) = \emptyset, ~~ \text{almost every} ~~ \omega.
$$
\end{thm}


\thebibliography{xx}

\bibitem{ma}
M.~Aizenman.
\newblock Localization at weak disorder: Some elementary bounds.
\newblock {\em Rev. Math. Phys.}, 6:1163--1182, 1994.

\bibitem{ag}
M.~Aizenman and S.~Graf.
\newblock Localization bounds for electron gas.
\newblock {\em Preprint mp\_arc 97-540}, 1997.

\bibitem{am}
M.~Aizenman and S.~Molchanov.
\newblock Localization at large disorder and at extreme energies: an elementary
  derivation.
\newblock {\em Commun. Math. Phys.}, 157:245--278, 1993.

\bibitem{ms1}
A.~Boutet de Monvel and J.~Sahbani.
\newblock On the spectral properties of discrete Schr\"odinger
operators.
\newblock {\em C. R. Acad. Sci. Paris} 326 Seri I; 1145-1150, 1998.

\bibitem{ms2}
A.~Boutet de Monvel and J.~Sahbani.
\newblock On the spectral properties of discrete Schr\"odinger
operators: multidimensional case.
\newblock {\em To appear in Rev. Math. Phys.} 

\bibitem{cl}
R.~Carmona and J.~Lacroix.
\newblock {\em Spectral theory of random {Schr\"odinger} operators}.
\newblock Birkh\"auser Verlag, Boston, 1990.

\bibitem{cfks}
H.~Cycon, R.~Froese, W.~Kirsch, and B.~Simon.
\newblock {\em Topics in the Theory of {Schr\"odinger} operators}.
\newblock Springer-Verlag, Berlin, Heidelberg, New York, 1987.

\bibitem{fp}
A.~Figotin and L.~Pastur.
\newblock {\em Spectral properties of disordered systems in the one body
  approximation}.
\newblock Springer-Verlag, Berlin, Heidelberg, New York, 1991.

\bibitem{kko}
W.~Kirsch, M.~Krishna and J.~Obermeit.
\newblock {Anderson} {Model} with decaying randomness-mobility
edge.
\newblock {to appear in Math. Zeit.}.

\bibitem{mk1}
M.~Krishna.
\newblock {Anderson model with decaying randomness - Extended
states.}
\newblock {\em Proc. Indian. Acad. Sci. ({Math}{Sci.})}, 100:220-240, 1990.

\bibitem{mk2}
M.~Krishna.
\newblock Absolutely continuous spectrum for sparse potentials.
\newblock {\em Proc. Indian. Acad. Sci. ({Math}{Sci.})}, 103(3):333--339, 1993.

\bibitem{ko}
M.~Krishna and J.~Obermeit.
\newblock Localization and mobility edge for sparsely random
potentials.
\newblock {Preprint xxx.lanl.gov/math-ph/9805015}.

\bibitem{jl2}
V.~Jaksic and Y.~Last.
\newblock Corrugated surfaces and A.C. spectrum 
\newblock {Preprint}

\bibitem{jl}
V.~Jaksic and Y.~Last.
\newblock Spectral properties of Anderson type operators 
\newblock {mp\_arc preprint 99-204}

\bibitem{jm}
V.~Jaksic and S.~Molchanov.
\newblock On the surface spectrum in Dimension Two  
\newblock {\em Helvetica Physica Acta}, 71: 169-183, 1999.

\bibitem{jm2}
V.~Jaksic and S.~Molchanov.
\newblock Localization of surface spectra 
\newblock {mp\_arc Preprint 99-196}

\bibitem{rs}
M.~Reed and B.~Simon.
\newblock {\em Methods of modern {Mathematical} {Physics}: {Functional}
  Analysis}.
\newblock Academic Press, New York, 1975.

\bibitem{bs}
B.~Simon.
\newblock Spectral analysis of rank one perturbations and applications.
\newblock In J.~Feldman, R.~Froese, and L.~Rosen, editors, {\em CRM Lecture
  Notes Vol. 8}, pages 109--149, Amer. Math. Soc., Providence, RI, 1995.

\bibitem{sw}
B.~Simon and T.~Wolff.
\newblock Singular continuous spectrum under rank one perturbations and
  localization for random {Hamiltonians}.
\newblock {\em Comm. Pure Appl. Math.}, 39:75--90, 1986.

\bibitem{st}
E.~Stein.
\newblock {\em Harmonic Analysis - Real variable methods,
Orthogonality and oscillatory integrals}
\newblock Princeton University Press, Princeton, New Jersey,
1993.

\bibitem{we}
J.~Weidman.
\newblock {\em Linear Operators in {Hilbert} spaces, GTM-68}.
\newblock Springer-Verlag, Berlin, 1987.

\endthebibliography
\end{document}